# High-Level Combined Deterministic and Pseudo-exhuastive Test Generation for RISC Processors

Adeboye Stephen Oyeniran
Department of Computer Systems
Tallinn University of Technology
Estonia
adeboye.oyeniran@taltech.ee

Raimund Ubar
Department of Computer Systems
Tallinn University of Technology
Estonia
raiub@pld.ttu.ee

Maksim Jenihhin
Department of Computer Systems
Tallinn University of Technology
Estonia
maksim@pld.ttu.ee

Cemil Cem Gürsoy
Department of Computer Systems
Tallinn University of Technology
Estonia
cemil@ati.ttu.ee

Jaan Raik
Department of Computer Systems
Tallinn University of Technology
Estonia
jaan@pld.ttu.ee

*Abstract*—Recent safety standards set stringent requirements for the target fault coverage in embedded microprocessors, with the objective to guarantee robustness and functional safety of the critical electronic systems. This motivates the need for improving the quality of test generation for microprocessors. A new high-level implementation-independent test generation method for RISC processors is proposed. The set of instructions of the processor is partitioned into groups. For each group, a dedicated test template is created, to be used for generating two test programs, for testing the control and the data paths respectively. For testing the control part, a novel high-level control fault model is proposed. Using this model, a set of deterministic test data operands are generated for each instruction of the given group. The advantage of the high-level fault model is that it covers larger than SAF fault class including multiple fault coverage in the control part. For generating the data path test, pseudo-exhaustive data operands are used. We investigated the feasibility of the approach and demonstrated high efficiency of the generated test programs for testing the execute module of the miniMIPS RISC processor.

## I. Introduction

Despite the fact that test generation for embedded processor cores of digital systems is a problem intensively investigated during decades in the test community, there is still a need for improvements in fault coverage and speed of test program generation in cases where no information about the details of implementation is given.

For the last decade, there has been an extensive research on Software-Based Self-Test (SBST) of processors [1-12]. The general idea of SBST is to use the resources of processors to test themselves, by running specific test programs. The nature of this method implies such features as non-intrusiveness, low cost and compatibility with at-speed and in-field testing [4-5]. SBST method is well accepted in industry. The interest in this method is growing in frames of in-field test for processor-centric systems in safety-critical applications [5-6]. Recent application domain standards, e.g. ISO26262, IEC61508, DO0254 set very stringent requirements for the target fault coverage in embedded microprocessor circuits, with the objective of guaranteeing robustness and functional safety of the critical electronic systems. Hence, more effort is being put into SBST for in-field test to satisfy these requirements. It is interesting to note at this point that one of the benefits of automated SBST is in reduction in test development cost [6-7].

SBST approaches can be structural and functional. Structural approaches [8-12], are based on test generation using information from lower level of design (gate-level or RTL-level description) of processors, whereas, functional approaches use mainly instruction set architecture (ISA) information. The structural approaches cannot be used when the structural information about the processors to be tested is not available. One of the first ISA based methods, using pseudo-random test sequences was proposed in [13]. Another solution, FRITS (Functional Random Instruction Testing at Speed) [14], was based on test program generation on random instruction sequences with pseudo-random data. It suits well for wafer test due to its cache-resident nature. Alternative cache-resident method for production testing [15] using random generation mechanism proves that high cost functional testers can be replaced by the low-cost SBST without significant loss in fault coverage. Another approach, based on evolutionary technique was proposed in [16]. Test program is being composed of the most effective code snippets (in a question of SAF coverage), which were distinguished by constant re-evaluation. The method, however, is based on structural information.

Later research concentrates on test approaches for specific processor parts like pipeline, branch prediction mechanism [17-18] or caches [19-20]. In [21], a method is proposed, which can enhance SBST program in order to bring more coverage to pipeline logic and also memory addressing. Another approach for testing the pipeline was made in [22]. The proposed strategy involves the activation of faults related to the data hazards and register forwarding logic in processor core, and later research concentrates on decode stage of the pipeline [5]. A variation of on-line SBST with the objective of enhancing lifetime reliability was proposed in [31].

In this paper, we propose a novel deterministic high-level test generation method for SBST of embedded processors which is based on a novel implementation-free high-level functional fault model. The advantage of the model is higher fault class than the well measurable standard single SAF, covering as well bridging and multiple SAF faults in the control part. The determinism of the fault model stands in a novel proposed set of data constraints to be satisfied by generating data operands to be used with instructions under test. For testing the data-path, pseudo-exhaustive data operands are used. Experimental result shows that the data constraints proposed for the control test contributes also noticeably to reaching high SAF coverage for the data-path test.

The rest of the paper is organized as follows. In section 2, we present a novel high-level control fault model for microprocessors, and in section 3, we investigate the problem of mapping the high-level fault model to low gate-level faults. In section 4, we present a fault simulation algorithm, and discuss the problems of high-level fault coverage measurement. Section 5 is devoted to the overall composition of test programs. In section 6, we present experimental data, and section 7 concludes paper.

II. HIGH-LEVEL CONTROL FAULT MODEL FOR PROCESSORS

The purpose of this research is to propose a novel method for testing RISC microprocessors in a functional way and without resorting to the knowledge of implementation details.

The main concept of the proposed method is based on partitioning the set of instructions of the processor under test into groups which can be tested by test templates which includes initialization, instruction under test, and observation of the results, in a similar way as in [5]. In this paper, we focus on testing of the executing units in pipelined RISC processors consisting of a control part and data path as shown in Fig.1. The method can be generalized also for testing other specific parts of microprocessors, such as other pipeline stages, register decoding, flag testing, branch prediction mechanism etc.

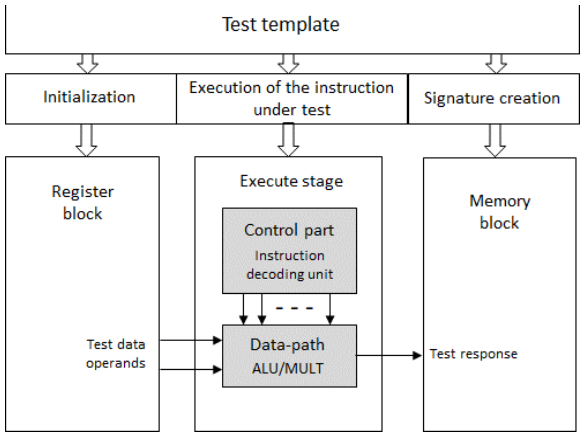

*Fig.1. Test execution set up*

The gray part of Fig.1 presents the test target which is the goal of this research. In Fig.2, we represent the execute unit in an implementation-free generic way as an equivalent circuit where the control part is highlighted as AND-OR multiplexer for decoding the instructions and extracting the results of the executed instructions. The circuit in Fig.2 represents equivalent disjunctive normal form (EDNF) related to the execute unit. The independence from implementation details results from the fact that a test developed for detecting all non-redundant faults in the EDNF, will also detect all faults in the original circuit [27]. Moreover, the exhaustiveness of the control signals together with the functional data constraints as the basis of the proposed method will target larger fault class than traditionally measured single SAF coverage contributes.

Assume, the ALU executes $n$ different functions $y = f_i(d)$ by a set $F = \{f_i\}$ of instructions, where $d$ represents data operand(s) for $f_i$, where the length of the data word (operand) is $m$, and ALU is controlled by $p$ control signals. In Fig.1, the control part consists of the multiplexer MUX and $p$ control lines (originating in the opcode field of the instruction register) as control inputs to MUX. The $n$ AND blocks (consisting of $m$ AND gates) in the control part of the execute unit have each $p$ control and a single $m$-bit data input, whereas the OR block has $n$ data word inputs from the outputs of AND blocks. Each AND block consists of $m$ AND gates with $p$ control inputs, and a single bit data input.

Let us classify two types of high-level functional fault models for the ALU: *control faults* (the faults related to the control part of the ALU), and *data faults* (the faults related to the data part of the ALU). For the control faults, we will introduce a novel high-level functional control fault model as follows.

Denote by $y_i$ the data word considered as the result of execution of the function $f_i$ with data operand(s) $d_i$ as $y_i = f_i(d_i)$.

**Definition 1.** Introduce for the function (instruction) $f_i \in F$, the following *high-level control fault model* $M(f_i)$ as a set of data operands $M(f_i) = \{D_i\}$, which satisfy the following constraints at least once for each bit $k$ of $y_i$:

$$\forall k \in (1,m): \{\exists d_i \in M(f_i) \ (y_{i/k} \neq 0)\}, \quad (1)$$

$$\forall f_j \in F, j \neq i : \forall k \in (1,m)\{\exists d_i \in M(f_i) \ (y_{i/k} < y_{j/k})\} \quad (2)$$

Depending on the technology, implemented in the microprocessor, the constant 0 in formula (1) can be changed into 1, and instead of the relation "<" in formula (2), there can be ">".

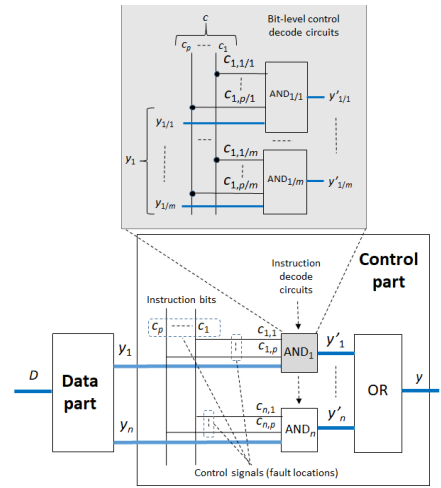

*Fig.2. Generic DNF based control structure of ALU*

The constraint (1) is needed for testing that the function $f_i$ can be executed and the result "$y_i = 1$" can be produced in each bit of the data word to detect the faults SAF/0 on all inputs of AND-gates. The constraint (2) is needed for testing that the result "$y_i = 0$" can be produced in each bit of the data word to decect two types of faults: SAF/1 on all inputs of the AND-gates related to the function $f_i$, and all functional faults of overwriting the value "$y_i = 0$" in each bit due to the control faults of other functions $f_j$, $j \neq i$.

The proposed fault model can be regarded as a generalization of the *conditional SAF model* or *input pattern fault model* (similar to ones considered in [23-26]). In case of conditional SAF, we are testing SAF on the gate-level lines at some constrained signals on other lines, whereas in case of the proposed high-level fault model of Definition 1, we are testing the instructions of microprocessors at a set of constraints for data (operands).

There are two novelties of this approach. First, due to using the EDNF based (not optimized) control unit model, the generated test may be over dimensioned. Second, the functional constraints (1) and (2) tend to produce more test patterns than it is needed for only single SAF detection. However, both aspects work in favour of larger fault class coverage, including multiple faults also, as already mentioned.

The size (complexity) of the proposed high-level control fault model can be represented by the number of data constraints to be satisfied, that is $C = n(n-1)mp$

## III. MAPPING OF HIGH-LEVEL FAULTS TO GATE-LEVEL FAULTS

Introduce the following notations of the input information for solving the problem.

**Definition 2.** Let $D^*_i$ be the set of data operands which satisfy the constraints of the fault model $M(f_i)$, $T^*_i$ is the test for the instruction $f_i$, which uses the data operands $d \in D^*_i$, and $T^* = \{T^*_i\}$ is the full test, generated for all high-level control faults for the set of instructions $F = \{f_i\}$.

**Theorem 1.** The test $T^* = \{T^*_i\}$, which covers all non-redundant high-level faults of the fault model $M(f_i)$, covers also all gate-level non-redundant SAF in the control part of the microprocessor, which controls the set of functions $F$.

Proof. The proof can be done in 2 steps. Firstly, consider the equivalent circuit of ALU control part presented in Fig.2, and described as the following DNF

$$y = c_{1,1}c_{1,2}\ldots c_{1,p}y_1 \vee c_{2,1}c_{2,2}\ldots c_{2,p}y_2 \vee \ldots \vee c_{n,1}c_{n,2}\ldots c_{n,p}y_n \quad (3)$$

for each bit of the data word in the output of OR block. We can easily show that from generation of data which satisfy the constraints (1) and (2) for all functions $f_i \in F$, it follows that in the DNF all SAF faults will be detected. In this DNF the variables $c_{i,j}$ for selecting the data results $y_i$, $i = 1, \ldots n$, represent the global control signals $c_j$, $j = 1, \ldots p$, being either inverted or not, and covering in general case exhaustively all the $2^p$ combinations. Secondly, assume that the control circuit is optimized and is represented as a multi-level combinational circuit instead of the two-level DNF. In this case, we can represent the circuit as an equivalent disjunctive normal form in a similar way as DNF (3). As already mentioned, if there is a test set which detects all non-redundant faults in the EDNF, this test will detect also all faults in the original possibly optimized multi-level circuit [27]. ■

**Corollary 1.** If a high-level test is generated, so that the the constraints (1) and (2) are fully satisfied, but if there are some SAF in the related EDNF, which remain not detected by the high-level test, the not detected SAF are redundant.

**Corollary 2.** If there are some cases in the constraints (2), which cannot be satisfied by selecting data operands, these cases refer to the high-level redundancies in the model $M(f_i)$.

**Corollary 3.** If the high-level redundancies can be removed from $M(f_i)$, and the high-level test is generated, the not detected SAF are redundant.

**Example 1.** Consider a simplified ALU unit which implemets the set of three functions $f_1, f_2, f_3$, activated by a set of control signals $\bar{c}_2 c_1, c_2 \bar{c}_1, c_2 c_1$ respectively. The ALU can be represented by the DNF:

$$y = \bar{c}_2 c_1 y_1 \vee c_2 \bar{c}_1 y_2 \vee c_2 c_1 y_3. \quad (4)$$

The test $T^* = \{T^*_1, T^*_2, T^*_3\}$ generated for the control part of ALU that satisfies the constraints (2) is depicted in Table 1.

*Table 1. Example of a high-level control test*

| $T^*_i$ | Test | | | | | Fault table | | | Constraints satisfied |
|---|---|---|---|---|---|---|---|---|---|
| | $c_2$ $c_1$ | $y_1$ $y_2$ $y_3$ | | | | $\bar{c}_2 c_1 y_1$ | $c_2 \bar{c}_1 y_2$ | $c_2 c_1 y_3$ | |
| $T^*_1$ | 0 1 | 0 1 1 | | | | 1 1 **0** | 0 0 1 | **0** 1 1 | $y_1 < y_2$, $y_1 < y_3$ |
| $T^*_2$ | 1 0 | 1 0 1 | | | | 0 0 1 | 1 1 **0** | 1 **0** 1 | $y_2 < y_1$, $y_2 < y_3$ |
| $T^*_3$ | 1 1 | 1 1 0 | | | | **0** 1 1 | 1 **0** 1 | 1 1 **0** | $y_3 < y_1$, $y_3 < y_2$ |

The table contains the test patterns in column 2, the fault table in columns 3-5, and the constraints satisfied by generating data for the control test patterns in column 6. The detected gate-level faults in the fault table are highlighted by red colour: 0 means the value of a signal which activates the fault SAF/1. For example, in case of the fault $c_2 \equiv 1$ in column 5, the value of the output signal $y = y_1 = 0$ will change from 0 to $y = y_1 \vee y_3 = 1$. For detecting the faults SAF/0, 3 more test patterns are needed (not shown in the table).

We see in the fault table that the faults $c_1 \equiv 1$ in column 3 and $c_2 \equiv 1$ in column 4 are not detected. Based on Corollary 1, these faults are redundant. By minimizing the function (4), we get a new formula

$$y = \bar{c}_2 y_1 \vee c_2(\bar{c}_1 y_2 \vee c_1 y_3).$$

where the redundancies are removed, and all SAF/1 are detectable by the test $T^*$.

The case of high-level redundancies is discussed in the following Sections.

Note, Theorem 1 and Corollaries 1-3 were formulated, considering the single SAF model. In fact, the power of the proposed high-level control fault model stretches far beyond the fault class of single SAF, as it will be shown in the following corollaries.

**Corollary 4.** The test $T^* = \{T^*_i\}$, covers all gate-level multiple SAF and bridging faults between control lines in the control part of the microprocessor, which controls the set of functions $F = \{f_i\}$.

Proof. From (2) it follows that for each function $f_i \in F$, $\forall k: (y_{i/k} < y_{j/k})$ for all $j \neq i$ must hold. This means that not only SAF/1 in a single control signal of a single function $f_j \in F$, $j \neq i$, can be detected (by overwriting $y_{i/k} = 0$ with $y_{j/k} = 1$), where the control words for $f_i$ and $f_j$ differ in a single bit, rather such overwriting of signals $y_{i/k} = 0$ with 1 can happen, and hence, can be detected, due to multiple changes $0 \rightarrow 1$ for $f_j \in F$, $j \neq i$, leading to detecting multiple faults. On the other hand, from the constraints (1-2), and from the exhaustiveness of testing all the control functions function $f_j \in F$, $j \neq i$, it follows that non-redundant bridging faults between the control lines can be also detected by $T^*$. ■

In case, when the target would be to detect only single SAF, then the fault model defined by the constraints (1) and (2) is over-dimensioned. For the case of full single SAF coverage, it would be sufficient to loosen the constraint (2) to

$$\forall f_j \in F, (HD(f_j, f_i) = 1), j \neq i:$$
$$\forall k \in (1, m) \{\exists d_i \in M(f_i) \ (y_{i/k} < y_{j/k})\},$$

where $HD(f_j, f_i) = 1$ is the constraint that the Hamming distance between the control codes for $f_j$ and $f_i$ must be 1. This simplication is similar to the approach used in [5]

The size of the reduced high-level control fault model applied only to the code-neighboring functions $f_j$, $f_i$ with $HD(f_j, f_i) = 1$, is equal to $C_{red} = nmp < C = n(n-1)p$.

## IV. HIGH-LEVEL FAULT COVERAGE

To measure the fault coverage for the fault model $M(f_i)$, $f_i \in F$, proposed in Definition 1, by the given test $T^*_i$ and the set of operands $D^*_i$, we introduce the high-level fault table as a matrix $E = ||e_{i,j}||$ with $n$ columns and $n$ rows, where $n$ – is the number of functions in $F$. Each entry $e_{i,j}$ in $E$ is a $m$-bit vector $e_{i,j} = (e_{i,j/1}, e_{i,j/2}, \ldots, e_{i,j/m},)$, where $m$ is the number of bits in the data-words $y_i = f_i(d_i)$, $d_i \in D^*_i$. We denote by $e_{i,j/k} = 1$, if the constraint $y_{i/k} < y_{j/k}$ for the bit $k$ in the set of constraints (2) is satisfied by the set of data operands in $D^*_i = \{d_i\}$, and $e_{i,j/k} = 0$ if not.

*Table 2. Example of a High-Level Fault Table*

|       | $f_1$ - MOV | $f_2$ - ADD | $f_3$ - SUB | $f_4$ - CMP | $f_5$ - AND |
|-------|-------------|-------------|-------------|-------------|-------------|
| $f_1$ - MOV |   | 111111 | 111111 | 111111 | **000000** |
| $f_2$ - ADD | 11111 |   | 11111**0** | 111111 | 111111 |
| $f_3$ - SUB | 11111 | 11111**0** |   | 111111 | 111111 |
| $f_4$ - CMP | 111111 | 111111 | 111111 |   | **000000** |
| $f_5$ - AND | 11111 | 111111 | 111111 | 111111 |   |

An example of the matrix $E = ||e_{i,j}||$ for a test $T^*$ for a set of functions $F = \{f_i\}$ executed by the set of instructions $I = \{MOV, ADD, SUB, CMP, AND\}$, is presented in Table 2. Each $i$-th row in the table represents the high-level control fault coverage of testing the function $f_i \in F$, (and the respective instruction $I_i \in I$.

The fault table $E = ||e_{i,j}||$ is the result of high-level fault simulation for the given set of operands $D^*_i$, to be used by the high-level test $T^*_i$. In this paper we have implemented the following high-level control fault simulation algorithm.

**Algorithm 1.**
(1) **for** all row instructions $f_i$, $i = 1, \ldots, n$
(2)    **for** all data operands $d_{i,j,1}, d_{i,j,2}, j = 1, \ldots, n_i$
(3)       **for** all column instructions $f_h$, $h = 1, \ldots, n$
(4)          calculate the value $y_h$
(5)          check the relation $y_i < y_h$, $h \neq i$
(5)          update the vector $e_{i,h} \in E$
(6)       **end for** column instructions
(7)    **end for** data operands
(8) **end for** row instructions

Based on Algorithm 1, we implemented a simulation based high-level test generation method on the basis of random search for test data to satisfy the constraints (2).

In Table 2, 0s refer either to not detected high-level control faults or to the possible high-level redundancies of the faults related to the constraints $y_{i/k} < y_{j/k}$, where $i$ and $j$ correspond to the rows and columns, respectively, and $k$ refers to the bit number. All 0s in $e_{ij}$ refer to high probability of the redundancy of the high-level fault model.

In most cases of ALU operations (like for $e_{15}$ and $e_{45}$ in Table 2), it is very easy to identify this type of redundancy. For example, if $y_i = f_i(a, b)$ refers to the AND operation and $y_j = f_j(a, b)$ refers to OR, it is straightforward that the constraint $y_i < y_j$, i.e. $(a \lor b) < (a \land b)$ cannot be satisfied by any values for $a$ and $b$.

In cases when there is an entry $e_{i,j/k} = 1$ in a single bit $k$ of the vector $e_{ij}$ (like for $e_{23}$ and $e_{32}$ in Table 2), or in only few bits of the vector $e_{ij}$, we can suggest for the redundancy proof a method called "*partial truth table method*". The idea of the method stands in showing the equivalence of partial truth tables (or to prove the impossibility of solving the related constraints) for the functions involved in the constraint relation, so that as few as possible responsible bits should be selected for the need of the proof.

*Table 3. Examples of high-level fault redundancy proofs*

| # | $y_{i/k} < y_{j/k}$ | $e_{ij}$ | $y_{i/k} < y_{j/k}$ | 00 | 01 | 10 | 11 |
|---|---------------------|----------|---------------------|----|----|----|----|
| 1 | SUB < ADD | 1…11**0** | SUB | 0 | 1 | 1 | 0 |
|   |           |           | ADD | 0 | 1 | 1 | 0 |
| 2 | OR < ADD  | 1…11**0** | OR  | 0 | 1 | 1 | 1 |
|   |           |           | ADD | 0 | 1 | 1 | 0 |
| 3 | OR < AND  | 0…000     | OR  | 0 | 1 | 1 | 1 |
|   |           |           | AND | 0 | 0 | 0 | 1 |
| 4 | OR < XOR  | 0…000     | OR  | 0 | 1 | 1 | 1 |
|   |           |           | XOR | 0 | 1 | 1 | 0 |

In Table 3, examples for 1-bit partial truth tables for the functions SUB, ADD, OR, AND, and XOR, for selected bits $k$ (shown with red color) are shown. The pairs 00, 01, 10, 11 in the title row represent the values of the data variables $d_{i/k}$ (as arguments for $y_{i/k}$) in bit $k$. The 1-bit values in the columns show the results of the related operations for the $k$-th bit. For the constraints SUB<ADD, and OR<ADD, the equivalence of the behavior in the least significant bit is demonstrated, which contradicts to the constraint (2). For the cases OR<AND, and OR<XOR, the missing of a solution for (2) is also shown for all possible input data combinations, and for all bits $k$. In some specific corner cases, the proof of redundancy may be more difficult.

The proof of high-level fault redundancy was not the target of the paper, and it needs special investigations. The quality of tests derived by the proposed method, SAF coverage was measured. The knowledge about redundancy of high-level faults is important when using of Corollary 3 for identification of redundant SAF by only applying fault simulation.

## V. HIGH-LEVEL TEST PROGRAM COMPOSITION

The full test $T$ for testing the set of functions $F = \{f_i\}$ can be represented as a set of subtests $T_i(f_i)$:

$$T = \{T_i(f_i)\} = \{(I_i, D_i) \mid i: f_i \in F\}$$

where $I_i$ denotes the instruction which executes the function $f_i \in F$, and $D_i$ denotes the set of data patterns (operands), each of them has to be used by the instruction $I_i$. The data patterns $d_{i,j} \in D_i$ may represent either single operands or concatenation of two operands $(d_{i,j,1}.d_{i,j,2})$ stored in the memory. For each group of similar instructions, there is a template – a subroutine, repeated in a loop for all instructions $I_i$, where $i : f_i \in F$, and each instruction $I_i$ is executed in a nested loop for all data operands in $D_i$, which are loaded by the initialization part of the template.

The architecture of test program is shown in Fig.3. The test tempates are created on the basis of Algorithm 2.

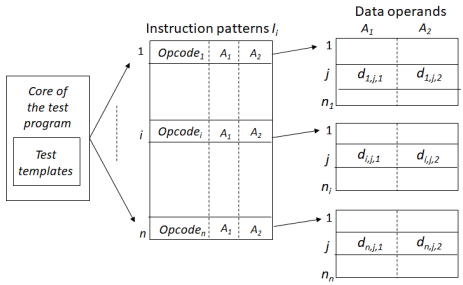

*Fig.3. Architecture of the test program*

**Algorithm 2.**

(1) **for** all instructions $I_i \in I$, $i : f_i \in F$
(2)   **for** all data operands $d_i \in D_i$
(3)     read $d_i$
(5)     execute the instruction $I_i$
(6)     store the test result $y_i = f_i(d_i)$
(7)   **end for** data
(8) **end for** instructions

Each subtest $T_i(f_i) \in T$ for testing $f_i \in F$ is partioned into two parts: test for the control part, and test for the data path. These two parts differ in how the data sets $D_i$ are generated.

For testing the control part, we use the data operands $D_i = D^*_i$, which are generated to satisfy the constraints of the fault model $M(f_i)$ according to Definition 1. For testing the data-path, for each instruction, dedicated data operands are to be generated. Denote these sets of operands as $D_i = D^{**}_i$.

Generation of the data operands to build the sets $D^{**}_i$ was not the objective of this paper. In the experimental research, to achieve the complete test results, we exploited for creating the data sets $D^{**}_i$ the parallel pseudoexhaustive test (PET) data operands, generated for selected data bits separately, and replicated then for other bits, using the methods presented in [28] for ALU, and in [29] for multiplication.

In this paper, we propose a new alternative approach for data-path testing, which directly results from the data operands generated for testing the control part – to execute each instruction using all data operands generated according to Definition 1 for all functions of the group $D^*$, so that

$$D_i = D^* = \cup_i D^*_i \mid i : f_i \in F.$$

In this data set, the data operands for testing the control and data paths are joined. This approach happened to be unexpectedly very efficient regarding the achieved SAF coverage, and at the same time, without adding cost for storing the test data in the memory.

Comparison of different approaches is presented in the Section for experiments.

## VI. EXPERIMENTAL RESULTS

We carried out experiments, consisting in high-level test data generation for the control and data parts of the execute stage of MiniMIPS processor [30], consisting of ALU and two multiplication modules MULT0 and MULT1.

The test program generation included automatic synthesis of test templates from manual parameter file, automated high-level test data (operands) generation to satisfy the constraints (1-2) and based on the fault simulation according to Procedure 1, and manual removal of the high-level fault redundancies to prove the 100% high-level test coverage.

To compare the quality of our high-level generated test program with commercial gate-level ATPG, we synthesized with Synopsys synthesis tool a gate-level implementation of the execute stage of MiniMIPS processor, and calculated with commercial fault simulation tool the gate-level SAF coverages for our high-level generated test program using two options of data sets described in Section V. The experimental research targeted 25 instructions $I_i \in I$ out of MiniMIPS 51 instructions, as the basis of the set of functions $F = \{f_i\}$ investigated in the paper.

Experimental results are shown in Table 4.

*Table 4. Experimental data*

| Quality measures | Parts of the execute module | # Faults | Proposed ATPG | | Gate level ATPG |
|---|---|---|---|---|---|
| | | | Only control data | Control + PET data | |
| Fault coverage % | Execute Stage | 203576 | 98.70 | 99.02 | 97.73 |
| | ALU | 2516 | 99.92 | 99.92 | 99.96 |
| | MULT0 | 95188 | 99.09 | 99.52 | 97.40 |
| | MULT1 | 91810 | 99.05 | 99.16 | 97.71 |
| # Stored test patterns | | | 166 | 166 | 957 |
| # Executed test patterns | | | 4150 | 4818 | 957 |
| Test generation time | | | 47s | Manually added PET data | 8h 27m |

We investigated two versions of test data generation. In the first version "only control data" we used the full data set $D^*$ generated automatically using the constraints (1-2). In the second version "control + PET data", we added to the data set $D^*$ additional manually generated pseudo-exhaustive test patterns, using the results in [29]. Both high-level tests were simulated by commercial tool to grade the gate-level SAF coverage. In both cases, the proposed method of high-level test generation, where the knowledge of implementation details was not needed, produced high gate-level SAF coverage for both, control and data parts of the execute module in MiniMIPS.

To evaluate the efficiency of the high-level ATPG, we used commercial gate-level ATPG for comparison. The time cost for high-level automated test generation is about two orders of magnitude less than the time cost of the commercial ATPG. The gate-level SAF coverages, achieved by the proposed method for the whole module under test, and also for the separate submodules ALU, MULT0 and MULT1 are significantly better than that of achieved by the commercial ATPG tool.

The proposed method has also advantage compared to the commercial gate-level ATPG in the number of test patterns to be stored in the memory. The test is stored in the compact form, unrolling only during the test execution.

## VII. CONCLUSIONS

In this paper, we proposed a new high-level test program generation method for execute modules of RISC microprocessors, which achieves gate-level SAF coverage significantly higher than a commercial gate-level ATPG. Furthermore, the speed of test generation exceeds the speed of the commercial ATPG more than two orders of magnitude.

The proposed method is based on a new high-level control fault model for microprocessors, which consists of a set of data constraints to be satisfied in test generation. The new test generation method uses as input information only the description of the instruction set, which is available in the manuals, and no knowledge of implementation details is needed.

The test is able to achieve very high coverage of non-redundant single SAF, as demonstrated by experiments.

Additional contribution of the paper, which shows advantage over state-of-the-art methods, is the coverage of a larger class of faults than only single SAF, including bridging faults and multiple SAF in the control parts under test. Hence, the proposed method for testing the control circuit faults is more powerful than the traditional gate-level ATPGs, which target only the single SAF fault class. However, this claim is based only on theoretical considerations. The related experimental research should be the future work.

The method was extended also to testing the faults in the data path of the execute modules of microprocessors. A metric of high-level fault coverage and a method for high-level fault simulation were developed. Additionally, a manual method for proof of high-level fault redundancies was also developed.

The future work will target optimization of test data operands, and the extensions of the proposed method for other modules of microprocessors not targeted in this paper.

ACKNOWLEDGMENT

The work has been supported in part by project H2020 MSCA ITN RESCUE (EU Horizon 2020, Grant 722325), Estonian research grant IUT 19-1 and Excellence Centre EXCITE in Estonia.

REFERENCES

[1] L.Chen, S.Dey. Software-based self-testing methodology for processor cores. IEEE Trans. on CAD of IC and systems, vol.20, no.3, 2001, pp. 369 - 380.

[2] N.Kranitis, A.Paschalis, D.Gizopoulos, G.Xenoulis. Software based self-testing of embedded processors. IEEE Trans. on Comp., vol.54, no.4, 2005.

[3] P.Bernardi, R.Cantoro, S.De Luca, E.Sanchez, A.Sansonetti. Development Flow for On-Line Core Self-Test of Automotive Microcontrollers. IEEE Trans. on Comp., v.65, no.3, 2016, pp-744-754.

[4] M.Psarakis, D.Gizopoulos, E.Sanchez, M.S.Reorda Microprocessor software-based self-testing. IEEE Design Test of Computers, v.27, no.3, 2010.

[5] P.Bernardi, R.Cantoro, L.Ciganda, E.Sanchez, M.S.Reorda, S.D.Luca, R.Meregalli, A.Sansonetti. On the in-field functional testing of decode units in pipelined risc processors. IEEE Int Symp. on Defect and Fault Tolerance in VLSI and Nanotechnology Systems. 2014, pp. 299–304.

[6] A.Riefert, R.Cantoro, M.Sauer, M.S.Reorda, B.Becker. A flexible framework for the automatic generation of SBST programs. IEEE Trans on VLSI Systems, vol.24, no.10, 2016, pp. 3055–3066.

[7] M.Schölzel, T.Koal, S.Rieder, H.T.Vierhaus. Towards an automatic generation of diagnostic in-field sbst for processor components. LATW, 2013.

[8] S.Gurumurthy, S.Vasudevan, J.A. Abraham. Automatic generation of instruction sequences targeting hard-to-detect structural faults in a processor. IEEE International Test Conference, 2006.

[9] L.Lingappan, N. K. Jha. Satisfiability-based automatic test program generation and design for testability for microprocessors. IEEE Trans. on VLSI Systems, vol.15, no.5, pp. 518–530, 2007.

[10] C.H.Wen, L.-C.Wang, K.-T.Cheng. Simulation-based functional test generation for embedded processors. IEEE Trans. on Comp., vol.55, no.11, 2006.

[11] N. Kranitis, A. Paschalis, D. Gizopoulos, and G. Xenoulis. Software based self-testing of embedded processors. IEEE Trans. on Comp., vol.54, no.4, 2005.

[12] C.H.Chen, C.K.Wei, T.H.Lu, H.W.Gao. Software-based self testing with multiple-level abstractions for soft processor cores. IEEE Trans on VLSI Systems, vol.15, no.5, pp. 505–517, 2007.

[13] J.Shen, J.A.Abraham. Native mode functional test generation for processors with applications to self test and design validation. Int. Test Conference, 1998..

[14] P.Parvathala, K.Maneparambil, W.Lindsay. Frits - a microprocessor functional bist method. International Test Conference, 2002, pp. 590–598.

[15] I.Bayraktaroglu, J.Hunt, D. Watkins. Cache resident functional microprocessor testing: Avoiding high speed io issues. IEEE Int. Test Conference, 2006.

[16] F.Corno, E.Sanchez, M.S.Reorda, G.Squillero. Automatic test program generation: a case study. IEEE Design Test of Computers, vol.21, no.2, 2004.

[17] D.Changdao, M.Graziano,E.Sanchez, M.Sonza Reorda, M. Zamboni, N. Zhifan. On the functional test of the BTB logic in pipelined and superscalar processors. LATW, 2013.

[18] E. Sanchez and M. S. Reorda. On the functional test of branch prediction units. IEEE Trans. on VLSI Systems, vol.23, no.9, 2015, pp. 1675–1688.

[19] S. D. Carlo, P. Prinetto, and A. Savino. Software-based self-test of setassociative cache memories. IEEE Trans. on Computers, vol.60, no.7, 2011, pp. 1030–1044.

[20] J.Perez Acle, R.Cantoro, E.Sanchez, M.Sonza Reorda. On the functional test of the cache coherency logic in multi-core systems. LATS, 2015.

[21] D.Gizopoulos, M.Psarakis, M.Hatzimihail, M.Maniatakos, A.Paschalis, S. Ravi, A.Raghunathan. Systematic software-based self-test for pipelined processors. IEEE Trans. on VLSI Systems, vol.16, no.11, 2008, pp.1441–1453.

[22] P.Bernardi, R.Cantoro, L.Ciganda, B.Du, E.Sanchez, M.S.Reorda, M.Grosso, O.Ballan On the functional test of the register forwarding and pipeline interlocking unit in pipelined processors. 14th Int. Workshop on Microprocessor Test and Verification, Dec 2013, pp. 52–57.

[23] K.B.Keller. Hierarchical Pattern Faults for Describing Logic Circuit Failure Mechanisms. US Patent 5546408, Aug. 13, 1994.

[24] R.Ubar. Fault Diagnosis in Combinational Circuits by Solving Boolean Differential Equations. Automation and Remote Control, Vol.40, No 11, part 2, Nov. 1980, Plenum Publishing Corporation, USA, pp. 1693-1703.

[25] R.D.Blanton, J.P.Hayes. On the Properties of the Input Pattern Fault Model. ACM Trans. Des. Automat. Electron. Syst., Vol. 8, No. 1, pp. 108-124, Jan. 2003.

[26] S.Holst, H.-J.Wunderlich. Adaptive Debug and Diagnosis Without Fault Dictionaries. Proc. of 13th ETS, Verbania, Italy, May 2008, pp.199-204.

[27] D.B.Armstrong. On Finding a Nearly Minimal Set of Fault Detection Tests for Combinational Logic Nets. IEEE Trans. on Electronic Computers, v.EC-15, no.1,1966 pp.66-73.

[28] A.S.Oyeniran, A.Jasnetski, A.Tsertov, R.Ubar. High-Level Test Data Generation for Software Based Self-Test in Microprocessors. 6th Mediterranean Conference on Embedded Computing (MECO 2017), 2017.

[29] A.S.Oyeniran, S.P.Azad, R.Ubar. Parallel Pseudo-Exhaustive Testing of Array Multipliers with Data-Controlled Segmentation. Int. Symp. on Circuits and Systems (ISCAS), 2018.

[30] OpenCores, "MiniMIPS ISA".

[31] F. Pellerey et al., "Rejuvenation of NBTI-Impacted Processors Using Evolutionary Generation of Assembler Programs," 2016 IEEE 25th Asian Test Symposium (ATS), Hiroshima, 2016, pp. 304-309